\documentclass[preprint,amsmath,amssymb,aip,apl]{revtex4-1}
\usepackage{graphicx}% Include figure files
\usepackage{dcolumn}% Align table columns on decimal point
\usepackage{bm}% bold math
\usepackage{float}
\begin{document}

\preprint{APS/123-QED}

\title{Electrical transport and low-temperature scanning tunneling microscopy \\of microsoldered graphene}% Force line breaks with \\

%\author{V.\ Geringer}
%\affiliation{ II. Institute of Physics, RWTH Aachen University and JARA-FIT, Otto-Blumenthal-Stra{\ss}e, 52074 Aachen}
%\author{T.\ Echtermeyer}
%\affiliation{Advanced Microelectronic Center Aachen (AMICA), AMO GmbH, Otto-Blumenthal-Stra{\ss}e 25, 52074 Aachen}
%\author{R.\ R{\"u}ckamp}
%\affiliation{ II. Institute of Physics, RWTH Aachen University and JARA-FIT, Otto-Blumenthal-Stra{\ss}e, 52074 Aachen}
%\author{M.\ Schmidt}
%\affiliation{ II. Institute of Physics, RWTH Aachen University and JARA-FIT, Otto-Blumenthal-Stra{\ss}e, 52074 Aachen}
%\author{M.\ Lemme}
%\affiliation{Advanced Microelectronic Center Aachen (AMICA), AMO GmbH, Otto-Blumenthal-Stra{\ss}e 25, 52074 Aachen}
%\author{M.\ Morgenstern}
%\affiliation{ II. Institute of Physics, RWTH Aachen University and JARA-FIT, Otto-Blumenthal-Stra{\ss}e, 52074 Aachen}

\author{V.~Geringer$^{1,3}$, D.~Subramaniam$^{1,3}$, A.~K.~Michel$^{1,3}$, B.~Szafranek$^2$, D.~Schall$^2$, A.~Georgi$^{1,3}$, T.~Mashoff$^{1,3}$, D.~Neumaier$^2$, M.~Liebmann$^{1,3}$ and M.~Morgenstern$^{1,3}$}
\affiliation{ $^{1}$II. Institute of Physics, RWTH Aachen
University,\\ Otto-Blumenthal-Stra{\ss}e, 52074 Aachen\\
$^{2}$Advanced Microelectronic Center Aachen (AMICA), AMO GmbH,\\
Otto-Blumenthal-Stra{\ss}e 25, 52074 Aachen, $^{3}$ JARA:
Fundamentals of Future Information Technology}

\date{\today}% It is always \today, today,

\begin{abstract}
Using the recently developed technique of microsoldering, we perform
systematic transport studies of the influence of PMMA on graphene
revealing a doping effect of up to $\Delta$n =
3.8$\times$10$^{12}$\,cm$^{-2}$, but negligible influence on
mobility and hysteresis. Moreover, we show that microsoldered
graphene is free of contamination and exhibits very similar
intrinsic rippling as found for lithographically contacted flakes.
Finally, we demonstrate a current induced closing of the previously
found phonon gap appearing in scanning tunneling spectroscopy,
strongly non-linear features at higher bias probably caused by
vibrations of the flake and a B-field induced double peak attributed
to the 0.Landau level.
\end{abstract}
\maketitle

The discovery of graphene in 2004 \cite{Novoselov04,Zhang05} with
its exceptional room-temperature mobility and its unconventional
Quantum Hall effect boosted a wealth of theoretical \cite{Neto09}
and experimental \cite{Geim07} work. However, several basic
properties like the morphology of the flakes
\cite{Meyer07,Geringer09} or the limiting factors of its mobility
\cite{Katsnelson08,Hwang07} are not settled. Since both of them
appear to depend on details of the preparation process, it is
crucial to investigate well defined samples. Here, we use the
recently developed technique of microsoldering
\cite{Girit07,Tapaszto08} in order to avoid the dirt usually induced
by lithography. Indeed, in contrast to lithographically contacted
samples, the microsoldered graphene is free of contamination as
evidenced by scanning tunneling microscopy (STM). Thus, we could
probe the influence of PMMA revealing that PMMA leads to a
considerable $n$-doping up to 3.8$\times$10$^{12}$\,cm$^{-2}$, while
mobility and voltage induced hysteresis are barely changed. Using
the clean sample, which exhibits intrinsic corrugation
\cite{Geringer09}, we investigated the local spectroscopic
properties of the flake at $T=5$ K. We show that the phonon-induced
gap found recently by Zhang et al. \cite{Zhang08} is closed at
higher tunneling current probably due to a local heating of the
sample. Additional features appear in scanning tunneling
spectroscopy (STS) at higher bias. They are not related to the local
density of states (LDOS), but are most likely induced by strongly
non-linear vibrations. Finally we observe a B-field induced feature,
which we attribute to the 0.Landau level.

\begin{figure*}[tp]
\centering
\includegraphics[width=\linewidth]{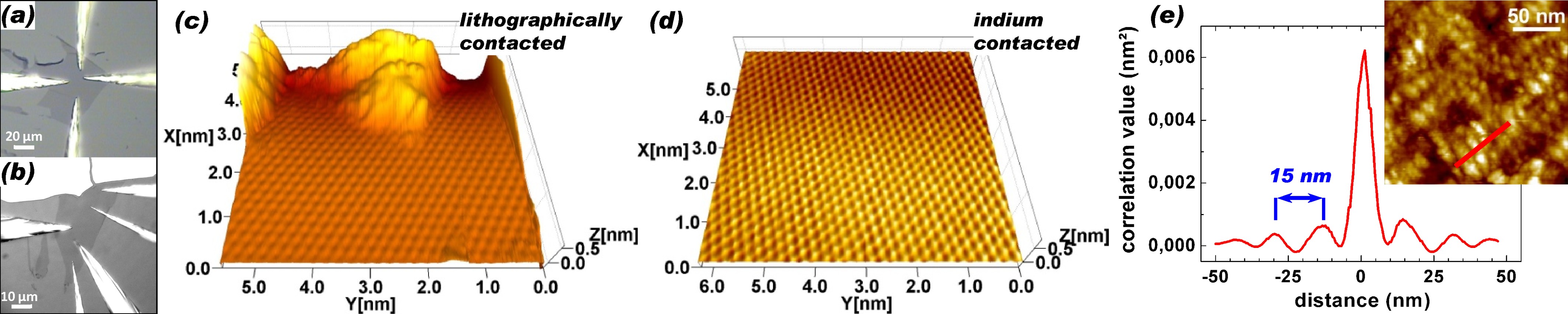}[tp]
\caption{\label{fig1} (Color online) (a), (b) Optical images of
microsoldered graphene flakes with four (a) and five (b) indium
contacts. (c) STM image of a lithographically contacted graphene
monolayer on Si/SiO$_2$ (T=5\,K, tunneling current I$_T$=0.2\,nA,
tunneling voltage U$_T$=1\,V); remaining dirt (probably PMMA) is
visible in the top part of the image. (d) STM image of a
microsoldered monolayer of graphene  on Si/SiO$_2$ (T=5\,K,
I$_T$=0.5\,nA, U$_T$=0.5\,V). (e) Inset: large-scale STM image of
microsoldered graphene (T=5\,K, I$_T$=1\,nA, U$_T$=0.5\,V); main:
correlation function averaged along the direction indicated as a
solid line in the inset; dominating wavelength is marked.}
\end{figure*}

Fig.~\ref{fig1}(a) and (b) show optical images of two microsoldered
monolayers of graphene, which are prepared under ambient conditions
by mechanical exfoliation on a 90\,nm SiO$_2$ layer on Si(001),
identified by Raman spectroscopy and soldered by drawing liquid
indium on top of the flake using a micromanipulator \cite{Girit07}.
The majority of the resulting contacts exhibits Ohmic behavior with
contact resistances of 1$-$50\,k$\Omega$. Fig.~\ref{fig1}(c) and (d)
show STM images of a graphene flake contacted by standard electron
beam lithography and lift-off \cite{Geringer09} (c) and by
microsoldering (d). They are recorded at $T=5$\,K in ultra-high
vacuum and are slightly high-pass filtered in order to suppress the
rippling, thereby increasing the visibility of the atomic
resolution. The lithographically contacted flake exhibits clusters
of dirt with heights up to 2 nm which indicate the remaining resist.
In contrast, the microsoldered sample is free of contamination. We
found that about 10 \% of the flake regions far away from the
contacts and 30 \% of the regions close to the contact are covered
with clusters of dirt after lithography, but we never found such
contamination on microsoldered samples imaging several $\mu$m$^2$ in
each case. The inset of Fig.~\ref{fig1}(e) shows a larger scale
image of the microsoldered graphene exhibiting an intrinsic rippling
of amplitude $A=1$\,nm  very similar to the rippling found on
lithographically contacted samples \cite{Geringer09, Deshpande09}.
The correlation function (main image) is taken along the main
direction of rippling (line in inset) being 30$^\circ$ with respect
to the C$-$C bond direction. The rippling again shows a preferential
wavelength of 15\,nm evidencing that intrinsic rippling does not
depend on the contact procedure.

\begin{figure}[tp]
\includegraphics{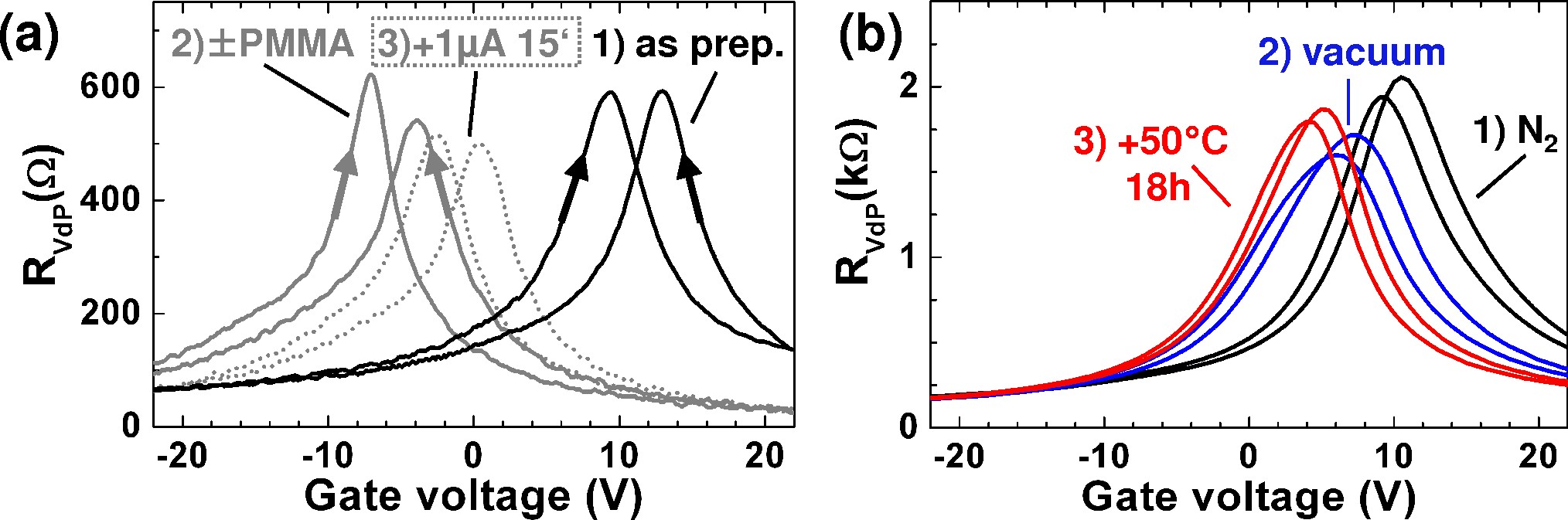}
\caption{\label{fig2} (Color online) (a) Four-terminal van der Pauw
measurement of the microsoldered graphene monolayer shown in
Fig.~\ref{fig1}(a) (ambient conditions); sweep rate:  0.5 V/s; black
curve: as-prepared sample, grey curve: same sample after coverage
with PMMA and subsequent cleaning, dotted grey curve: same sample
after additional current-induced heating by 1\,$\mu$A; measurement
direction is marked by arrows. (b) Four-terminal resistance of
microsoldered graphene in nitrogen (right curves), vacuum (middle
curves) and after additional vacuum annealing at 50$\,^\circ$C for
18 h (left curves).}
\end{figure}

The gate voltage dependent 4-point resistance (T = 295\,K) of a
microsoldered sample is shown in Fig.~\ref{fig2}(a). The black curve
measured directly after microsoldering shows a hysteresis $\Delta
V_{\rm Gate}=4$\,V and a Dirac point at V$_{\rm Dirac}$ = 9\,V/13\,V
depending on sweep direction. Using five different samples, we get a
mobility of $\mu$ = 3200\,$\pm$\,600\,cm$^2$/Vs from van-der-Pauw
measurements in agreement with two-point measurements of
\cite{Girit07}. In vacuum (p = 2$\times$ 10$^{-5}$\,mbar), $\mu$
increases by 50\,\%  after 18 h at T = 50\,$^\circ$C. The hysteresis
$\Delta V_{\rm Gate}$ strongly depends on
 gate voltage range, sweep rate and surrounding atmosphere.
For the sake of comparison, all curves in Fig.~\ref{fig2} are
measured with very similar parameters as used for lithographically
contacted samples in \cite{Lohmann09}. In ambient conditions, we
find $V_{\rm Dirac} = 3-9$\,V during downward sweep and $\Delta
V_{\rm Gate} = 3-4$\,V. While $\Delta V_{\rm Gate}$ is very similar
to the results of \cite{Lohmann09} (after correcting for different
SiO$_2$ thickness), $V_{\rm Dirac}$ shows much less scatter. By
changing the environment to pure nitrogen and vacuum, $V_{\rm
Dirac}$ and $\Delta V_{\rm Gate}$ are continuously reduced down to
$V_{\rm Dirac } = 2$\,V and $\Delta V_{\rm Gate} = 0.5$\,V (Fig.
\ref{fig2}(b)). This improvement is well known also for
lithographically contacted samples and usually attributed to the
removing of water and an according reduction of charge trapping
\cite{Lohmann09, Kim03, Wang09}.

Since the main difference between microsoldered and lithographically
contacted samples is the scatter of $V_{\rm Dirac}$, we investigate
the influence of PMMA, probably the major contamination after
lithography, in more detail. The three curves in Fig.~\ref{fig2}(a)
show the same sample directly after microsoldering (black), after
covering the surface with PMMA using a solution in anisol and a
subsequent standard cleaning procedure of rinsing the sample in
acetone and propanol (grey full line), and after additional heating
by currents of 1\,$\mu$A for several minutes (dotted grey
line)\cite{Moser07}. The curves recorded after PMMA contamination
but without cleaning look very similar to the ones with cleaning.
Obviously neither $\Delta V_{Gate}$ nor the mobility (steepness of
curves) are strongly influenced by PMMA, but a significant n-type
doping results, which is straightforwardly deduced to be $\Delta$n =
3.8$\times$10$^{12}$\,cm$^{-2}$. Moreover, the resistance curve is
more asymmetric after contamination and, thus, comparable to most
measurements of lithographically contacted samples \cite{Lohmann09,
Lemme07, Chen08, Tan07}. We attribute the reduction of n-doping by
current heating (dotted curve) to removing of solvents. Further
heating by 10\,$\mu$A does not shift $V_{\rm Dirac}$ anymore and
$V_{\rm Dirac}$ can be shifted reversibly between the two points
(grey curves) by repeated PMMA contamination and current heating.
Thus, the PMMA process leads to two kinds of n-dopants, one,
probably the solvents, being removed by moderate heating. On the
basis of our results, we explain the n-doping of graphene partially
observed after lithography \cite{Lohmann09, Lemme07, Chen08, Tan07}
by residues of PMMA. Thereby, we resolve the puzzle that the known
p-doping by O$_2$, H$_2$O \cite{Schedin07, Leenaerts08, Wehling08}
or by Au-contacts \cite{Giovannetti08, Lee08} cannot explain the
occasionally observed negative Dirac point positions. Notice, that
PMMA has recently also been found to bury p-dopants, which then
cannot be removed by additional heating \cite{Lohmann09}.
\begin{figure*}[tp]
\centering
\includegraphics[width=\linewidth]{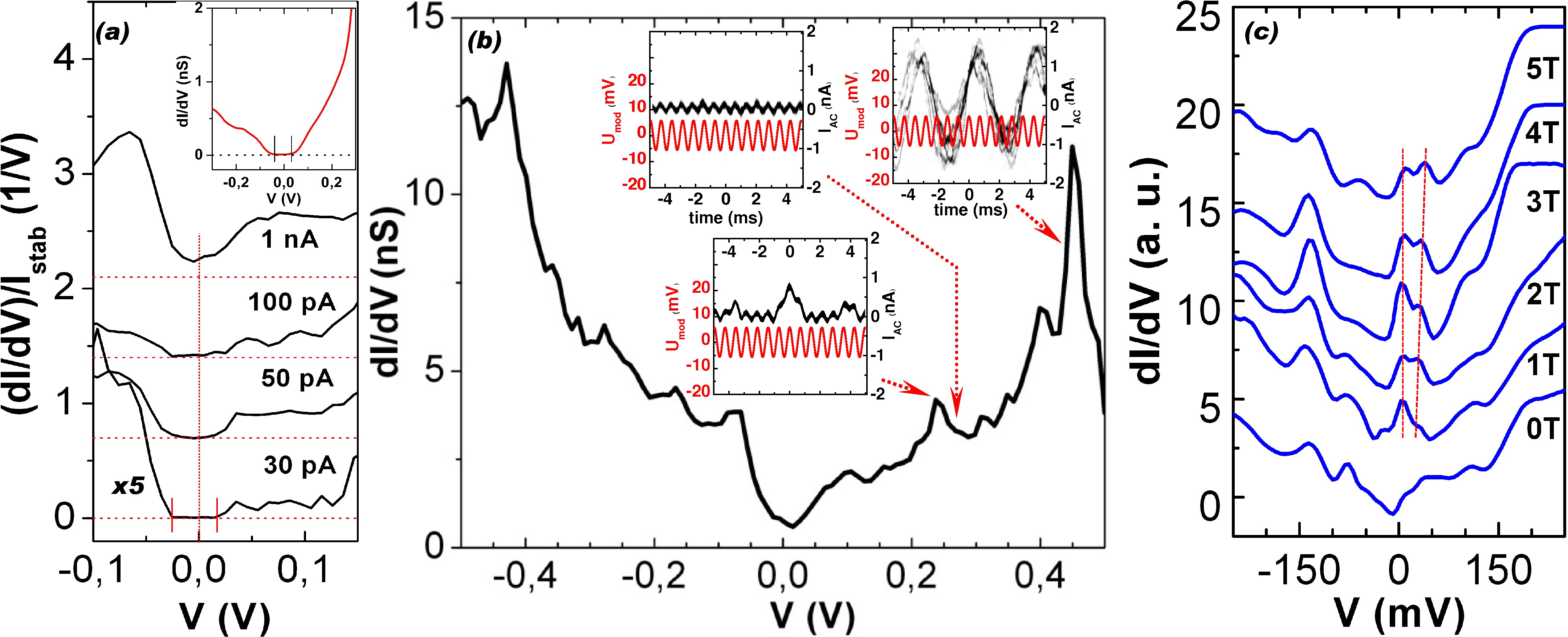}
\caption{\label{fig3} (Color online) (a) $dI/dV$ spectra (normalized
by stabilization current $I_{\rm stab}$) taken at the same position
with the same tip for different $I_{\rm stab}$ as marked (T=5\,K,
V$_{stab}$=0.5\,V, V$_{mod}$=4\,mV); curves are vertically displaced
for clarity; dotted, horizontal lines mark $dI/dV/I_{\rm stab}=0$/V.
Inset: dI/dV spectrum of lithographically contacted graphene
(T=5\,K, I$_{stab}$=2\,nA, V$_{stab}$=0.7\,V, V$_{mod}$=1\,mV);
vertical lines mark the phonon gap. (b) dI/dV spectrum taken on
microsoldered graphene at higher $I_{\rm stab}$ (T=5\,K,
I$_{stab}$=2\,nA, V$_{stab}$=0.5\,V, V$_{mod}$=4\,mV); insets show
current response I$_{AC}$ (black) to the oscillating tip voltage
$U_{\rm mod}$ (red) at the DC voltages marked by arrows. (c)
$B$-field dependence of $dI/dV$ spectra (T=5\,K, I$_{stab}$=2\,nA,
V$_{stab}$=0.07\,V, V$_{mod}$=10\,mV); curves are vertically
displaced for clarity; dashed lines mark the splitting of the
0.\,LL.}
\end{figure*}

Finally, we discuss STS results from the microsoldered sample.
Fig.~\ref{fig3}(a) shows a series of dI/dV spectra recorded with the
same microtip at the same position. At low current, a gap of $\Delta
V \simeq 40$\,mV is visible, being very similar to the gap found by
Zhang et al. \cite{Zhang08}. The gap was interpreted as a phonon
gap, i.e. a phonon with large wave vector is required in order to
tunnel into or out of K-point states. Such a gap has not been found
by other groups \cite{Tapaszto08, Deshpande09, Luican09} and we also
found it only occasionally using lithographic samples (inset of
Fig.~\ref{fig3}(a)). Figure \ref{fig3}(a) demonstrates that the gap
disappears at larger current. It reappears after reducing the
current again (not shown). The onset current for closing the gap
obviously depends on the microtip. This straightforwardly explains
the discrepancy found by different groups. We believe that a local
heating of the sample by the tunneling current produces enough
phonons, so that the phonon annihilation can provide the required
wave vector towards K-point states. Fig.~\ref{fig3}(b) shows a
larger scale $dI/dV$ curve ( high $I_{\rm stab}$). It exhibits a
number of peaks. The peaks are not related to the density of states
as can be deduced from the current response to the applied
modulation voltage shown as insets. At the peaks, the frequency of
the response differs strongly from the excitation frequency. Since
this effect is reproducible on the same position, appears at
different voltages for different positions and has never been
observed on Au(111) with the same setup, we attribute it to the
properties of the graphene sample. We suggest that dielectric forces
are responsible, which might lead to a non-linear mechanical
movement of the flake \cite{Mashoff09}. Thus, dI/dV spectroscopy on
graphene seems to be very susceptible to signals not related to the
LDOS.

Fig.~\ref{fig3}(c) shows a B-field run of spectra, which does not
feature series of Landau levels, although we have measured Landau
and spin levels on InSb(110) with the same STM \cite{Mashoff08,
Becker10}. Only around 0\,V, there is a peak, which develops into a
doublet at higher $B$. It looks similar to the 0.\,Landau level
observed by STS of flat graphene either on HOPG \cite{Luican09} or
on the C-face of SiC(0001) \cite{Miller09}. The splitting of the
doublet of 25\,meV at B=4\,T corresponds exactly to the splitting
values found in \cite{Luican09} and \cite{Miller09}.

In summary, we used microsoldering in order to show that the
mobility, the hysteresis and the rippling of graphene on SiO$_2$ are
barely influenced by lithography, but that PMMA and solvents lead to
significant n-type doping. Moreover, we confirm a considerable
improvement of cleanliness by microsoldering, which is a big
advantage for future STM experiments. First STS results show that
$dI/dV$ curves are strongly susceptible to effects not related to
the LDOS probably due to local heating and mechanical vibrations.
Nevertheless, B-field data reveal a peak which we relate to the
split 0.\,Landau level.

We appreciate helpful discussions with C. Stampfer and acknowledge
financial support by the DFG (Mo-858/8-1 and Mo 858/11-1) and the
excellence initiative of the German federal and state government.

%\bibliography{Literatur}% Produces the bibliography via BibTeX.

\end{document}